\documentclass[letterpaper,12pt]{article}

\usepackage{endnotes}
\usepackage{amssymb}
\usepackage{amsfonts}
\usepackage{amsmath}
\usepackage{amsthm}
\usepackage{graphicx,color}
\usepackage{comment}

\usepackage{array}

\usepackage{url}

\usepackage{setspace}
\def\R{\mathbb{R}}
\def\endproof{\hfill\diamondsuit}

\def\sF{{\mathcal F}}

\def\sA{{\mathcal A}}

\def\L{\mathbb{L}}

\def\E{\mathbb{E}}

\def\sF{\mathcal{F}}
\def\P{\mathbb{P}}

\def\Q{\mathbb{Q}}

\numberwithin{equation}{section}
\theoremstyle{plain}                
\newtheorem{theorem}{Theorem}[section]
\newtheorem{lemma}[theorem]{Lemma}

\theoremstyle{definition}           
\newtheorem{definition}[theorem]{Definition}

\newtheorem{assumption}[theorem]{Assumption}
\theoremstyle{remark}               

\usepackage[margin=1.10in]{geometry}

\begin{document}
\pagenumbering{arabic} \pagestyle{plain}

\begin{center}
\LARGE{\bf Radner equilibrium in incomplete L\'evy models}\footnote{The first author has been supported by the National Science Foundation under Grant No. DMS-1411809 (2014-2017). Any opinions, findings, and conclusions or recommendations expressed in this material are those of the author(s) and do not necessarily reflect the views of the National Science Foundation (NSF). 
}
\end{center}
\begin{center}
\ \\

{\large \bf Kasper Larsen}\\ Department of Mathematical Sciences, \\
  Carnegie Mellon University,\\ Pittsburgh, PA 15213, USA \\ email: {\tt
    kasperl@andrew.cmu.edu}

\ \\ 

{\large \bf Tanawit Sae Sue}\\ Department of Mathematical Sciences, \\
  Carnegie Mellon University,\\ Pittsburgh, PA 15213, USA \\ email: {\tt
    tsaesue@andrew.cmu.edu}

\end{center}
\begin{center}

{\normalsize \today }
\end{center}
\vspace{.5cm}

\begin{verse}
{\sc Abstract}: We construct continuous-time equilibrium models based on a finite number of exponential utility investors. The investors' income rates as well as the stock's dividend rate are governed by discontinuous L\'evy processes. Our main result provides the equilibrium (i.e., bond and stock price dynamics) in closed-form. As an application, we show that  the equilibrium Sharpe ratio can be increased and the equilibrium interest rate can be decreased (simultaneously) when the investors' income streams cannot be traded.

\end{verse}
\vspace{0.5cm}

\newpage

\section{Introduction}

We construct equilibrium models in which a finite number of heterogeneous exponential investors cannot fully trade their future income streams. We show that the framework of continuous-time L\'evy processes produces the Radner equilibrium in closed-form (i.e., optimal strategies, interest rates, drifts, and volatility structures are available in closed-form). Besides allowing for more model flexibility, we show that by going beyond models based on Brownian motions we can produce the following empirically desirable feature:  The class of pure jump L\'evy models can simultaneously lower the equilibrium interest rate and increase the equilibrium Sharpe ratio due to investors' income streams being unspanned (i.e., due to model incompleteness). 

The first construction of an incomplete continuous-time model which allows for an explicit description of the Radner equilibrium was given in \cite{CLM12}. As an application of this model, \cite{CLM12} show that model incompleteness can significantly lower the equilibrium interest rate. However, the (instantaneous) Sharpe ratio is unaffected by the model's incompleteness.\footnote{Theorem 4.1 in \cite{CL12} shows that no model based on exponential utilities, continuous consumption, and a filtration generated by Brownian motions can ever produce an incompleteness impact on the Sharpe ratio when this ratio is measured instantaneously.} Besides being of mathematical interest, our motivation behind extending the Brownian framework in \cite{CLM12} to the more general L\'evy framework is to produce simultaneously a negative impact on interest rate and a positive impact on the Sharpe ratio while still maintaining a closed-form equilibrium model. Our desire to construct an incomplete equilibrium model with these features is of course due to Weil's celebrated risk-free rate puzzle (see \cite{Wei1989}) as well as Mehra and  Prescott's equity premium puzzle  (see \cite{MP1985}). These and other asset pricing puzzles are also discussed in detail in the survey \cite{Cam00}.

The literature on continuous-time Radner equilibrium theory in models where the investors' income streams are spanned (i.e., complete models) is comprehensive and we refer to the recent references on endogenous completeness \cite{AR08}, \cite{HMT12}, \cite{HR13}, and \cite{Kra2015} for more information. On the other hand, models with continuous-time trading and unspanned income streams (i.e., incomplete models) are much less developed and only in recent years has progress been made. The papers \cite{Zit12}, \cite{Zha12}, \cite{CL15}, and \cite{KXZ15} consider models with exponential utilities, no dividends (i.e., only financial assets), and discrete-time consumption.\footnote{By restricting the investors to only consume at maturity, the economy's interest rate cannot be determined. Furthermore, by only considering financial assets, the assets' volatility structures also remain undetermined. Therefore, the interest rate and the volatility parameters are taken as exogenously specified model input in such models.} These papers differ in how general the underlying state-processes describing the investors' discrete-time income streams can be:  \cite{Zit12} considers a Brownian motion and an independent indicator process.  \cite{Zha12} and \cite{CL15} consider multiple Brownian motions (\cite{CL15} also allow for processes with mean reversion) whereas the recent paper \cite{KXZ15} allows for a non-Markovian Brownian setting. The current paper is more related to \cite{CLM12} and \cite{CL12} who - in Brownian settings - consider both financial and real assets in the case of exponential investors with continuous-time consumption. Indeed, the current paper can be seen as a direct extension of \cite{CLM12} to the setting of discontinuous L\'evy processes. 

The paper is organized as follows: The next section describes the underlying L\'evy framework. Section 3 provides the solution to the individual investors' problems. Section 4 contains our main result which provides the equilibrium parameters in closed-form. The last section illustrates numerically the equilibrium impacts due to incompleteness in the Gaussian compound Poisson case. The appendices contain all the proofs.

\section{Mathematical Setting}

\subsection{Underlying L\'evy process}
We let $T>0$ denote the time-horizon and we let $I$ denote the number of investors. $(\Omega,\sF,(\sF_t)_{t\in[0,T]},\P)$  denotes the underlying filtered probability space and we assume that $\sF = \sF_T$. 
For some underlying $\R^{I+1}$-dimensional pure jump L\'evy process $\eta$ we denote by $N = N(dt,dz)$ the random counting measure on $[0,T]\times \R^{I+1}$ associated with $\eta$'s jumps. The corresponding compensated random measure is denoted $\tilde{N}(dt,dz) = N(dt,dz) - \nu(dz)dt$ where $\nu$ is referred to as the L\'evy measure on $\R^{I+1}$ associated with $\eta$'s jumps, see, e.g.,  \cite{App2009} and \cite{Sat1999}  for more details about these objects. We assume that $(\sF_t)_{t\in[0,T]}$ is the filtration generated by $\eta$ (right-continuous and completed).

The following regularity assumption on the L\'evy measure $\nu$ will be made throughout the paper:

\begin{assumption}\label{ass:1} In addition to the usual properties 
\begin{align}\label{ass1:usual}
\nu(\{0\})=0,\quad \int_{\R^{I+1}}(||z||^2\land 1) \nu(dz)<\infty,
\end{align} 
the L\'evy measure $\nu$ satisfies the following three conditions:
\begin{align}
&\int_{||z||<1} |z^{(0)}| \nu(dz)<\infty,\label{ass1:smalljumps}\\
&\int_{||z||\ge1} e^{u^{(0)}z^{(0)} + u^{(i)}z^{(i)}} \nu(dz)<\infty \text{ for all } u^{(0)},u^{(i)}\in\R \text{ and }i=1,...,I,\label{ass1:bigjumps}\\
&\nu(z^{(0)}>0) > 0\text{ and } \nu(z^{(0)}<0) > 0.\label{ass1:pm} 
\end{align}
\end{assumption}
$\endproof$

Assumption \ref{ass:1} requires a few remarks: \cite{CLM12} consider the case of  correlated Brownian motions with drift which is why we focus exclusively on the pure jump case. The requirement that \eqref{ass1:bigjumps} holds for all $u^{(0)}$ and $u^{(i)}$ in $\R$ can be relaxed to a certain domain at the cost of more cumbersome notation (this can be seen from the proofs in Appendix B). Condition \eqref{ass1:smalljumps} is not implied by \eqref{ass1:usual}  because it requires that $\nu$ can integrate $z^{(0)}$ instead of $(z^{(0)})^2$ on the unit ball and has a number of implications; e.g., \eqref{ass1:smalljumps} 
ensures that the process 
\begin{align}\label{J}
J_t= \int_{||z||<1} z^{(0)} N(dz,dt), \quad t\in[0,T],
\end{align}
is well-defined and is of finite variation.\footnote{The process $J$ will be related to the stock's dividend process $D$ in the next section.} We note that $J$ can still  have infinite activity on finite intervals.  The last condition \eqref{ass1:pm} can also be relaxed to requiring that a certain explicit function is onto (see the last part of Lemma \ref{lem:varphi} in Appendix A).


\subsection{Exogenously specified model input}
The $I$ investors are assumed to have heterogeneous exponential utilities over running consumption, i.e., 
\begin{align}\label{Ui}
U_i(c) = -e^{-\frac{c}{\tau_i}},\quad c\in\R, \quad \tau_i>0, \quad i=1,...,I.
\end{align} 
Here the investor specific constants $(\tau_i)_{i=1}^I$ are referred to as the risk tolerance coefficients. 

We consider a pure exchange economy in the sense that bond prices, stock prices, income, and dividend processes are all quoted in terms of the model's single consumption good. The i'th investor's income rate process is modeled by
\begin{align}\label{dYi}
dY_{it} = \mu_idt + \sigma_i \int_{\R^{I+1}} z^{(i)} \tilde{N}(dt,dz), \quad Y_{i0} \in\R,
\end{align}
where $\mu_i$ and $\sigma_i> 0$ are constants for $i=1,...,I$. The single stock's dividend rate process is modeled by
\begin{align}\label{dD}
dD_t = \mu_Ddt + \sigma_D \int_{\R^{I+1}} z^{(0)} \tilde{N}(dt,dz), \quad D_0 \in\R,
\end{align}
where $\mu_D$ and $\sigma_D>0$ are constants. Because \eqref{J} is of finite variation, we see that $D$ is of finite variation too ($Y_i$ defined above by \eqref{dYi} might not be).

We note that the processes \eqref{dYi} and \eqref{dD} are not independent; indeed, the quadratic cross characteristics\footnote{The brackets $\langle\cdot ,\cdot\rangle$ are also called the conditional quadratic cross variation; see, e.g., Section III.5 in \cite{Pro04}.} between $D$ and $Y_i$ are given by
$$
d\langle Y_i, D\rangle_t =\int_{\R^{I+1}} \sigma_i\sigma_D z^{(0)}z^{(i)} \nu(dz),\quad \quad t\in[0,T], \quad i=1,...,I.
$$

\subsection{Endogenously determined price dynamics}
We will restrict the financial market to only consist of two traded securities (one financial asset and one real asset). The financial asset is taken to be the zero net supply money market account. Its price process will be shown to have the following equilibrium dynamics 
 \begin{align}\label{dS0}
dS^{(0)}_t = S^{(0)}_t r dt,\quad S_0^{(0)}=1,
\end{align}
where $r$ is a constant. Because the interest rate $r$ is deterministic, the money market account is equivalent to zero-coupon bonds of all maturities. In the following, we will need the corresponding annuity 
\begin{align}\label{def:A}
A(t) = \int_t^T e^{- r(s-t)}ds,\quad t\in[0,T].
\end{align}

The real asset is a stock  paying out the dividend stream $D$ (see \ref{dD}). This security is in unit net supply and we will show that its equilibrium price dynamics are given by 
 \begin{align}\label{dS}
dS_t +D_{t}dt = \Big(rS_{t} + \mu(t) \Big)dt + \sigma_DA(t) \int_{\R^{I+1}} z^{(0)} \tilde{N}(dt,dz),\quad S_{T-}=0,
\end{align}
for $t\in[0,T)$ where $\mu$ is a deterministic function. In order to have $S_t$ defined for $t\in [0,T]$, we explicitly define $S_T$ to be $S_{T-}$ (i.e., there is no jump in $S$ at $t=T$ by definition).

The following assumption is placed on the equilibrium output parameters and needs to be verified for any candidate set of parameters. 
\begin{assumption}\label{ass2} The function $\mu$ is continuous on $[0,T]$ and $\mu A^{-1}$ is constant. 
\end{assumption}
$\endproof$

As discussed in the Introduction we are interested in how model incompleteness impacts the interest rate $r$ and the stock's Sharpe ratio $\lambda$. The stock's (instantaneous) Sharpe ratio is defined as the constant 
\begin{align}\label{def:Sharpe}
\lambda= \frac{\mu(t)}{\sigma_DA(t)\sqrt{\int_{\R^{I+1}} (z^{(0)})^2 \nu(dz)}}.
\end{align}
The Sharpe ratio \eqref{def:Sharpe} measures the stock's return (cleaned for interest and dividend components) relatively to the standard deviation of its noise term. Sharpe ratios have been widely studied and used in the literature and we refer to \cite{CP2007} for an application of the Sharpe ratio \eqref{def:Sharpe} in a continuous-time jump diffusion setting. 

We end this section by introducing the set of equivalent martingale measures $\Q$ which will play a key role in what follows. For a process $\psi=\psi_t(z)$ we consider sigma-martingales $Z^\psi$ of the linear form
\begin{align}\label{Zpsi}
dZ^\psi_t = Z_{t-}^\psi \int_{\R^{I+1}} \psi_t(z) \tilde{N}(dt,dz),\quad t\in[0,T],\quad Z^\psi_0=1.
\end{align}
To ensure that a unique solution of \eqref{Zpsi} exists, we require that $\psi$ is $\tilde{N}$-integrable in the sense that $\psi$ is a predictable flow satisfying the integrability condition
\begin{align}\label{psi:c1}
\P\left( \int_0^T\int_{\R^{I+1}} \psi_t(z)^2 \nu(dz)dt <\infty\right)=1. 
\end{align}
To ensure that the solution of \eqref{Zpsi} is strictly positive we require
\begin{align}
&\psi_t(z) >-1,\,\P\text{-a.s.,}\,\text{  for all }z\in\R^{I+1} \text{ and } t\in[0,T].\label{psi:c2}
\end{align}
Under conditions \eqref{psi:c1} and \eqref{psi:c2}, the unique solution of \eqref{Zpsi} is a strictly positive local martingale. We additionally require that this process $Z^\psi$ is a martingale. This martingale property allows us to define the associated probability measure $\Q = \Q^\psi$ on $\sF_T$ by the Radon-Nikodym derivative $\frac{d\Q}{d\P}= Z_T^\psi$. The final requirement we place on the integrand $\psi$ is the sigma-martingale property under $\Q$ of the process $S_t/S^{(0)}_t + \int_0^t  D_u/S^{(0)}_u du$, $t\in[0,T]$. It\^o's product rule can be used to see that this requirement is equivalent  to the property\footnote{Conditions \eqref{ass1:usual} and \eqref{ass1:bigjumps} of Assumption \ref{ass:1} allow us to use Cauchy-Schwartz's inequality together with \eqref {psi:c1} to see that $z^{(0)}\psi_t(z)$ is $\nu$-integrable.}
\begin{align}\label{psi:c3}
\frac{\mu(t)}{\sigma_DA(t)} + \int_{\R^{I+1}} \psi_t(z) z^{(0)}\nu(dz) = 0,\, \P\text{-a.s.},\; \text{for all }t\in[0,T].
\end{align}
When $\psi$ satisfies the above requirements we refer to the associated measure $\Q=\Q^\psi$ as an equivalent martingale measure. Finally, we note that because the model $(S^{(0)},S)$ is incomplete, there exist infinitely many integrands $\psi$ satisfying the above requirements (Theorem \ref{thm:investor} and   Theorem \ref{thm:eq} below explicitly construct such integrands).

\section{Individual investors' optimization problems}
In this section the price dynamics \eqref{dS0} and \eqref{dS} as well as Assumption \ref{ass2} are taken as input and we consider the i'th investor's utility maximization problem. Because $S^{(0)}_0=1$, the investor's initial wealth is given by $X_{i0} = \theta_{i0-}^{(0)} + \theta_{i0-} S_0$ where investor's i'th initial endowments are $\theta_{i0-}^{(0)}$ units of the money market account and $\theta_{i0-}$ units of the stock.  In the following we will let $c_i$ denote the consumption rate in excess of the income rate $Y_i$, i.e., investor i'ths cumulative consumption at time $t\in[0,T]$ is given by $\int_0^t (c_{iu}+Y_{iu})du$.

We next describe the i'th investor's set of admissible strategies $\sA_i = \sA_i(\Q^{(i)})$ where $\Q^{(i)}$ is some investor specific equivalent martingale measure (Theorem \ref{thm:investor} below gives the specific $\Q^{(i)}$). The investor can choose predictable processes $\theta = (\theta_t)_{t\in[0,T]}$ and $c_i= (c_{it})_{t\in[0,T]}$
to generate the self-financing gain dynamics 
\begin{align}\label{dX}
dX_{it} = \Big(r X_{it} -c_{it} +\theta_t \mu(t)\Big) dt + \theta_t \sigma_DA(t) \int_{\R^{I+1}} z^{(0)} \tilde{N}(dt,dz),\quad X_{i0}\in\R,
\end{align}
provided that the various integrals exist on $[0,T)$ and provided that the left limit $X_{iT-}$ exists. In that case, we define the terminal value $X_{iT}=X_{iT-}$ in order to have $X_t$ defined for all $t\in[0,T]$. The investor is required to leave no financial obligations behind at maturity in the sense that
$$
 X_{iT} \ge0,\quad \P\text{-a.s.}
$$ 
Finally, to rule out arbitrage opportunities, we require the process $X_{it}/S^{(0)}_t + \int_0^t c_{iu}/S^{(0)}_udu$ is a $\Q^{(i)}$-supermartingale for $t\in[0,T]$. When these requirements are satisfied we write $(\theta,c_i)\in \sA_i$.

The investor's maximization problem is given by
\begin{align}\label{primal}
\sup_{(\theta,c_i)\in\sA_i}\E\left[\int_0^T U_i (c_{it} +Y_{it}) dt\right],
\end{align}
where the exponential utility function $U_i$ is defined by \eqref{Ui}. The rest of this section is devoted to describing the solution of \eqref{primal}.  To do this, we first note that  Lemma \ref{lem:varphi} in Appendix A ensures that the function
\begin{align}\label{fi}
\R\ni u^{(0)} \to \int_{\R^{I+1}} z^{(0)} e^{u^{(0)} z^{(0)} + u^{(i)}z^{(i)}} \nu(dz),\quad u^{(i)}\in\R,
\end{align}
has a well-defined continuous inverse $f^i_{u^{(i)}}(\cdot)$ with domain $\R$. We can then define the constants $\theta^*_i\in\R$ by (here we use \eqref{ass1:smalljumps} and the assumption that $\mu A^{-1}$ is constant)
\begin{align}\label{thetaistar}
\theta^*_i = -\frac{\tau_i}{\sigma_D}f^i_{-\tfrac{1}{\tau_i}\sigma_i}\left(-\frac{\mu(t)}{\sigma_DA(t)}+\int_{\R^{I+1}} z^{(0)}\nu(dz)\right),
\end{align}
where $A$ is the annuity defined by \eqref{def:A}. This allows us to define the constants $g_i$ by
\begin{align}
\begin{split}
g_i &= -r +\frac1{\tau_i}A(t)^{-1} \theta^*_i \mu(t) +\frac1{\tau_i}\mu_i\\
&- \int_{\R^{I+1}} \Big(e^{-\tfrac1{\tau_i}\theta^*_i \sigma_D z^{(0)} -\frac1{\tau_i} \sigma_i z^{(i)}}-1+\frac1{\tau_i}\theta^*_i \sigma_D z^{(0)}+\frac1{\tau_i} \sigma_i z^{(i)}\Big) \nu(dz).\label{go}
\end{split}
\end{align}
In terms of these objects, the following result provides the explicit solution to \eqref{primal}; see Appendix B for the proof. This shows that Theorem 1 in \cite{CLM12} carries over from the Brownian framework to the current setting of  L\'evy  processes. 

\begin{theorem}\label{thm:investor} Let Assumptions \ref{ass:1} and \ref{ass2} hold. Then the deterministic integrand $\psi_i$ defined by
\begin{align}\label{psii}
\psi_{i}(z) = e^{-\tfrac1{\tau_i}\theta^*_i \sigma_Dz^{(0)} -\tfrac1{\tau_i} \sigma_i z^{(i)}}-1,\quad z\in\R^{I+1},
\end{align}
produces an equivalent martingale measure $\Q^{(i)}$ via the martingale density \eqref{Zpsi}. Furthermore, for the corresponding admissible set $\sA_i= \sA_i(\Q^{(i)})$, the processes $(\theta^*_i,c^*_i)\in\sA_i$ attain the supremum in  \eqref{primal} where $\theta^*_i$ is defined by \eqref{thetaistar} and
\begin{align}\label{cistar}
c^*_{it} =  A(t)^{-1}X^*_{it} +\tau_i A(t)^{-1} \int_t^T e^{-r(s-t)} g_i(s-t)ds,\quad t\in[0,T].
\end{align}
In \eqref{cistar} the process  $X^*_i$ denotes the gain process \eqref{dX} produced by $(\theta^*_i,c^*_i)$.
\end{theorem}
The proof of Theorem \ref{thm:investor} establishes that the optimal controls $(\theta^*_i,c^*_i)$ produce a corresponding gain process $X^*_{i}$ with the property 
$$
X_{iT}^*=0,\quad \P\text{-a.s.},
$$
which means that optimally the investor leaves no wealth behind. 

\section{Radner equilibrium}
This section contains our main result which provides the Radner equilibrium in closed-form.  We start by defining what we mean by an equilibrium in the present setting:
\begin{definition}[Radner] We call $(S^{(0)},S)$ given by \eqref{dS0} and \eqref{dS} 
for an equilibrium if these price processes are produced by a pair $(r,\mu)$ satisfying Assumption \ref{ass2} and if:
\begin{enumerate}
\item There exist equivalent martingale measures $(\Q^{(i)})_{i=1}^I$ and processes $(\theta^*_i,c^*_i) \in \sA_i=\sA_i(\Q^{(i)})$ which attain the supremum in  \eqref{primal} for $i=1,...,I$. 
\item The markets clear in the sense that for all $(t,\omega)\in[0,T]\times\Omega$ we have
\begin{align}\label{clearingconds}
\sum_{i=1}^I c^*_{it} = D_t 
,\quad \sum_{i=1}^I \theta^*_{it} =1, \quad \sum_{i=1}^I \theta^{(0)*}_{it} =0.
\end{align}
\end{enumerate}
\end{definition}
$\endproof$

Our main existence result (the proof is in Appendix B) is stated in terms of two deterministic functions $(r,\mu)$: First we define the Sharpe ratio $\lambda\in\R$  (constant) through the requirement
\begin{align}\label{eq:mu0}
\sigma_D + \sum_{i=1}^I \tau_i f^i_{-\tfrac{1}{\tau_i}\sigma_i}\left(-\lambda\sqrt{\int_{\R^{I+1}} (z^{(0)})^2 \nu(dz)}+\int_{\R^{I+1}} z^{(0)}\nu(dz)\right)=0,
\end{align}
where the inverse functions $(f^i)_{i=1}^I$ are defined in the previous section (see \ref{fi}). Lemma \ref{lem:varphi} in Appendix A ensures that \eqref{eq:mu0} uniquely determines $\lambda$. Then we can define constants $\theta^*_i$ for $i=1,...,I$ by
\begin{align}\label{thetaistar1}
\theta^*_i= -\frac{\tau_i}{\sigma_D}f^i_{-\tfrac{1}{\tau_i}\sigma_i}\left(-\lambda\sqrt{\int_{\R^{I+1}} (z^{(0)})^2 \nu(dz)}+\int_{\R^{I+1}} z^{(0)}\nu(dz)\right).
\end{align}
In turn, this allows us to define the constant
\begin{align}
\begin{split}
r &= \frac1{\tau_\Sigma}\Big(\mu_D + \sum_{i=1}^I \mu_i - \int_{\R^{I+1}} \Big(\sum_{i=1}^I\tau_i e^{-\frac1{\tau_i}\theta^*_i\sigma_Dz^{(0)} -\frac1{\tau_i}\sigma_iz^{(i)}}-\tau_\Sigma +\sigma_Dz^{(0)}+\sum_{i=1}^I \sigma_iz^{(i)}\Big)\nu(dz)\Big),\label{eq:r}
\end{split}
\end{align}
where $\tau_\Sigma = \sum_{i=1}^I\tau_i$. Finally, we can define the annuity $A(t)$ by \eqref{def:A} for $r$ defined by \eqref{eq:r} and we can define the drift 
\begin{align}\label{eq:mu}
\mu(t) = \lambda\sqrt{\int_{\R^{I+1}} (z^{(0)})^2 \nu(dz)}\sigma_DA(t), \quad t\in[0,T].\end{align}

\begin{theorem}\label{thm:eq} Let Assumption \ref{ass:1} hold, let $\sum_{i=1}^I \theta_{i0-}^{(0)}=0$, $\sum_{i=1}^I \theta_{i0-}=1$,  and define the above  functions $(r,\mu)$ by \eqref{eq:r}-\eqref{eq:mu} as well as
\begin{align}\label{psi*}
\psi^*_t(z) = e^{f\left(-\frac{\mu(t)}{\sigma_DA(t)}+\int_{\R^{I+1}} z^{(0)}\nu(dz)\right)z^{(0)}}-1,\quad z\in\R^{I+1}, \quad t\in[0,T],
\end{align}
where $f$ is the inverse of the mapping
\begin{align}\label{f}
\R\ni u^{(0)} \to \int_{\R^{I+1}} z^{(0)}e^{u^{(0)} z^{(0)}}\nu(dz).
\end{align}
Then $(S^{(0)},S)$ with $S^{(0)}$ defined by \eqref{dS0} and $S$ defined by
\begin{align}\label{def:S}
S_t = \E^{\Q^*}\left[\int_t^T e^{-r(s-t)} D_s ds\Big|\sF_t\right],\quad t\in[0,T],
\end{align}
where $\Q^*$ is the equivalent martingale measure defined via $Z^{\psi^*}$, constitute an equilibrium.
\end{theorem}
We note that $\psi^*$ defined by \eqref{psi*} does not depend on time because \eqref{eq:mu} ensures that $\mu A^{-1}$ is constant. 

\section{Application}
In this section we will compare the incomplete equilibrium of Theorem \ref{thm:eq} with the corresponding complete equilibrium based on the representative agent. In the second part of this section we specify the  L\'evy measure $\nu$ to be the widely used compound Poisson process with Gaussian jumps and illustrate numerically the impacts on the resulting parameters due to model incompleteness.\footnote{The geometric form of this L\'evy process was first used in finance by Merton in his classical paper \cite{Mer76}. It is also the basis for Bates' asset pricing model developed in \cite{Bat1996}.}

\subsection{Representative agent's equilibrium}
It is well-known that when all investors have exponential utilities, then so does the sup-convolution describing the representative agent's preferences with risk tolerance coefficient $\tau_\Sigma = \sum_{i=1}^I\tau_i$; see, e.g., Section 5.26 in \cite{HL88}. We therefore define the representative agent's utility function by
\begin{align}\label{Urep}
U_{\text{rep}}(c) = -e ^{-c/\tau_\Sigma},\quad c\in\R.
\end{align}

The consumption-based capital asset pricing model developed in \cite{Bre1979} (and extended in \cite{GS1982} to certain incomplete models) is based on constructing price processes by applying the first-order condition for optimality in the representative agent's problem through the proportionality requirement
\begin{align}\label{naiveFOC}
U'_{\text{rep}}(D_t + \sum_{i=1}^I Y_{it}) \propto 
e^{-r^{\text{rep}}t} Z_t^\text{rep},\quad t\in[0,T].
\end{align}
Here $r^\text{rep}$ is the interest rate and $Z^{\text{rep}}$ is the model's (unique) martingale density. This model (i.e., $r^\text{rep}$ and   $Z^{\text{rep}}$) will serve as the basis for our comparison. It\^o's lemma produces the following dynamics of the left-hand-side of \eqref{naiveFOC} 
\begin{align*}
&\frac{dU'_{\text{rep}}(D_t + \sum_{i=1}^I Y_{it})}{U'_{\text{rep}}(D_t + \sum_{i=1}^I Y_{it}) }\\&= \int_{\R^{I+1}} \big(e^{-\tfrac1{\tau_\Sigma}\big(\sigma_Dz^{(0)} +\sum_{i=1}^I\sigma_iz^{(i)}\big)}-1\big) \tilde{N}(dt,dz)-\frac1{\tau_\Sigma}\Big(\mu_D +\sum_{i=1}^I\mu_i\Big)dt \\&+ \int_{\R^{I+1}} \Big(e^{-\tfrac1{\tau_\Sigma}\big(\sigma_Dz^{(0)} +\sum_{i=1}^I\sigma_iz^{(i)}\big)}-1+\frac1{\tau_\Sigma}\big(\sigma_Dz^{(0)} +\sum_{i=1}^I\sigma_iz^{(i)}\big)\Big) \nu(dz)dt.
\end{align*}
By matching coefficients with the right-hand-side of \eqref{naiveFOC} we find
\begin{align}
dZ_t^{\text{rep}} &= Z_{t-}^{\text{rep}}\int_{\R^{N+1}}\psi^\text{rep}(z)\tilde{N}(dt,dz), \quad \psi^\text{rep}(z) = e^{-\tfrac1{\tau_\Sigma}\left(\sum_{i=1}^I\sigma_iz^{(i)}+\sigma_D z^{(0)}\right)}-1,\\
r^{\text{rep}} &=\frac1{\tau_\Sigma}\Big(\mu_D +\sum_{i=1}^I\mu_i \Big)\label{rrep}\\&- \int_{\R^{I+1}} \Big(e^{-\tfrac1{\tau_\Sigma}\big(\sigma_Dz^{(0)} +\sum_{i=1}^I\sigma_iz^{(i)}\big)}-1+\frac1{\tau_\Sigma}\big(\sigma_Dz^{(0)} +\sum_{i=1}^I\sigma_iz^{(i)}\big)\Big) \nu(dz).\nonumber
\end{align}
We denote by $\Q^{\text{rep}}$ the measure defined by $\frac{d\Q^{\text{rep}}}{d\P} = Z_T^\text{rep}$  on $\sF_T$. We find the parameters describing the stock dynamics by computing the dynamics of
\begin{align}\label{Srep}
S^\text{rep}_t = \E^{\Q^\text{rep}}\left[\int_t^T e^{-r^\text{rep}(s-t)}D_s ds\Big|\sF_t\right],\quad t\in[0,T].
\end{align}
As in the proof of Theorem \ref{thm:eq} in Appendix B, we find that the dynamics of \eqref{Srep} are of the form \eqref{dS} but with $r=r^\text{rep}, A=A^\text{rep}$ and $\mu = \mu^\text{rep}$ where 
\begin{align}
 A^\text{rep}(t)&= \int_t^T e^{-r^\text{rep}(s-t)}ds,\\
\mu^\text{rep}(t) &=- \sigma_DA^\text{rep}(t)\int_{\R^{I+1}} \psi^\text{rep}(z) z^{(0)}\nu(dz),
\end{align}
for $t\in[0,T]$. Finally, the Sharpe ratio based on the representative agent is defined as:
\begin{align}\label{def:Sharpe_rep}
\lambda^\text{rep}= \frac{\mu^\text{rep}(t)}{\sigma_DA^\text{rep}(t)\sqrt{\int_{\R^{I+1}} (z^{(0)})^2 \nu(dz)}},
\end{align}
which is the analogue of \eqref{def:Sharpe}.

\subsection{Incompleteness impacts in a numerical example}
In this section we consider the L\'evy measure corresponding to a compound Poisson process with Gaussian jumps (i.i.d. zero-mean normals with covariance matrix $\Sigma$) and a unit constant Poisson intensity. In other words, for a symmetric positive definite matrix $\Sigma$ with unit diagonal elements, we consider the L\'evy measure
\begin{align}\label{ex:compound}
\nu(dz) = \frac1{\sqrt{(2\pi)^{I+1}\text{det}(\Sigma)}}e^{-\frac12 z'\Sigma^{-1}z}dz,\quad z\in\R^{I+1}.
\end{align}
This measure satisfies Assumption \ref{ass:1}. Furthermore, the functions $f^i$ and $f$ (the inverse functions of \ref{fi} and \ref{f}) can be expressed via the Gaussian moment generating function $e^{\frac12 \eta'\Sigma\eta}$, $\eta\in\R^{I+1}$, and its derivatives.

Based on \eqref{eq:r} and \eqref{rrep}, we see that the incompleteness impact on the equilibrium interest rate is given by
\begin{align*}
r^{\text{rep}}-r
&=\int_{\R^{I+1}} \Big(\sum_{i=1}^I\frac{\tau_i}{\tau_\Sigma} e^{-\frac1{\tau_i}\theta^*_i\sigma_Dz^{(0)} -\frac1{\tau_i}\sigma_iz^{(i)}}-e^{-\tfrac1{\tau_\Sigma}\big(\sigma_Dz^{(0)} +\sum_{i=1}^I\sigma_iz^{(i)}\big)}\Big) \nu(dz).
\end{align*}
Jensen's inequality and the clearing property $\sum_{i=1}^I \theta^*_i=1$ (see \ref{eq:mu}) can be used to see that this difference is always non-negative (a similar observation is made in \cite{CLM12} and \cite{CL12}). On the other hand, the impact on the instantaneous Sharpe ratio due to model incompleteness, i.e., $\lambda - \lambda^\text{rep}$, can be both positive and negative. Here $\lambda^\text{rep}$ is defined by \eqref{def:Sharpe_rep} and the (instantaneous) Sharpe ratio $\lambda$ in the incomplete equilibrium is defined by \eqref{def:Sharpe} and is found implicitly by solving 
 \begin{align}\label{eq:mu1}
\sigma_D + \sum_{i=1}^I \tau_i f^i_{-\tfrac{1}{\tau_i}\sigma_i}\left(-\lambda
\right)=0.
\end{align}
This follows from \eqref{eq:mu0} and the zero-mean and unit variance properties of $\nu(dz)$.

\begin{figure}[!h]
\begin{center}
$\begin{array}{cc}
\includegraphics[width=7cm, height=5cm]{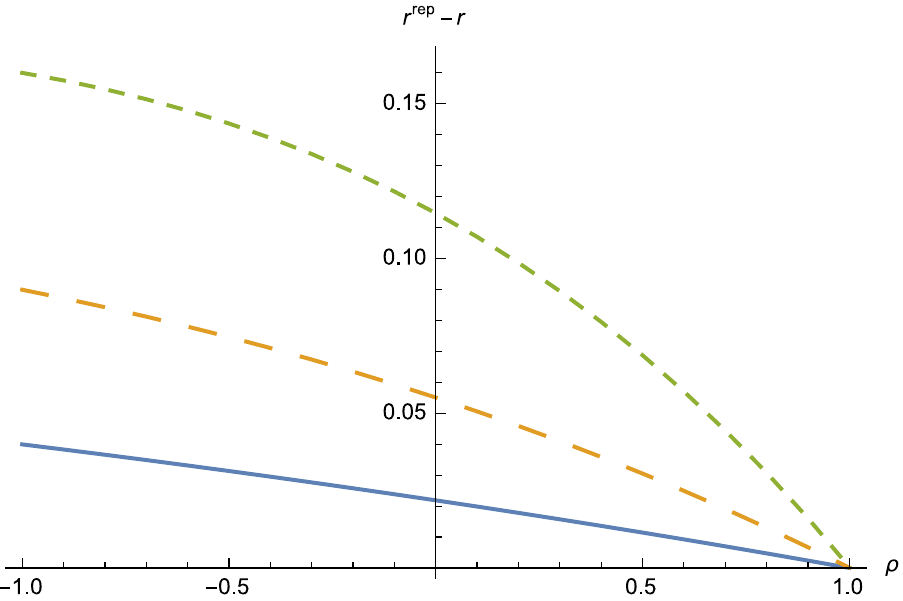} &
\includegraphics[width=7cm, height=5cm]{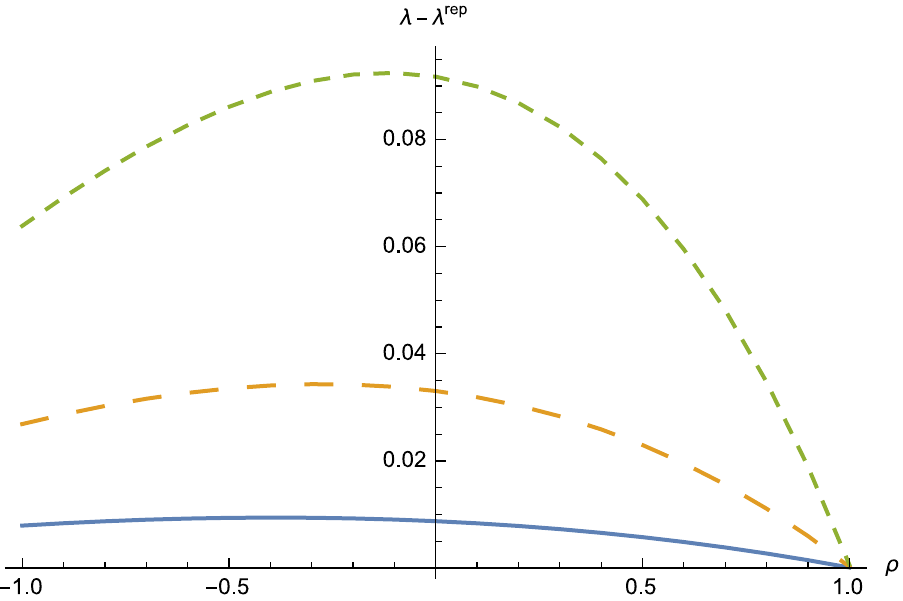}
\end{array}$
\caption{Plot of impacts due to model incompleteness on $r^\text{rep}-r$ (left) and $\lambda - \lambda^\text{rep}$ (right) seen as a function of the correlation coefficient $\rho$. We consider the limiting economy ($I\to\infty$) whereas the remaining parameters are given by \eqref{parameters} for the various risk tolerance coefficients $\tau$:  $\tau = \tfrac12$ (--------), $\;\tau = \tfrac13\;$ (-- -- --), and $\tau = \tfrac14$ (- - - -).  }
\label{fig1}
\end{center}
\end{figure}

To proceed with the numerics, we will use a flat correlation matrix in the sense that $\Sigma_{ij} = \rho$ for $i\neq j$ and $\Sigma_{ii}=1$ for $i,j=0,1,...,I$ where $\rho\in(-1,1)$.  The remaining parameters used to generate Figure \ref{fig1} are
\begin{align}\label{parameters}
\sigma_D = .2I, \quad \sigma_i=.1,\quad \tau_i = \tau, \quad i=1,...,I.
\end{align} 
\newpage

From Figure \ref{fig1} we see that our model simultaneously can produce a positive impact on the equilibrium (instantaneous) Sharpe ratio and a negative impact on the equilibrium interest rate. As discussed in the Introduction, these effects are empirically desirable because they are linked to the asset pricing puzzles in \cite{Wei1989} and \cite{MP1985}. Finally, we note that as $\rho\uparrow 1$ the resulting model approaches the representative agent's complete model and both incompleteness impacts vanish.

\appendix

\section{Two auxiliary lemmas }

\begin{lemma}\label{lem:varphi}
Suppose Assumptions \ref{ass:1} holds. Then the partial derivative 
\begin{align}\label{def:varphi}
\varphi_i(u^{(0)},u^{(i)}) = \frac{\partial }{\partial u^{(0)}} \int_{\R^{I+1}} e^{u^{(0)}z^{(0)} + u^{(i)}z^{(i)}} \nu(dz),\quad u^{(0)},u^{(i)}\in\R, 
\end{align}
is a well-defined function and satisfies the following properties:
\begin{enumerate}
\item The function $\varphi_i$ has the representation
\begin{align}\label{rep:varphi}
\varphi_i(u^{(0)},u^{(i)}) =  \int_{\R^{I+1}} z^{(0)}e^{u^{(0)}z^{(0)} + u^{(i)}z^{(i)}} \nu(dz),\quad u^{(0)},u^{(i)}\in\R.
\end{align}
\item The function $\varphi_i$ is jointly continuous.
\item For fixed $u^{(i)}\in\R$, the function $u^{(0)} \to \varphi_i(u^{(0)},u^{(i)})$ is strictly increasing and onto $\R$. Consequently, the inverse function $f^{(i)}_{u^{(i)}}(\cdot)$ exists and is continuous on $\R$.
\end{enumerate}
\end{lemma}
\proof For the first claim, we can use the bound
\begin{align}\label{bound1}
\frac{|e^{hz}-1|}{|h|}\le |z|e^{|z|},\quad z^{(0)}\in\R,\quad |h|\le1.
\end{align}
This bound is integrable by \eqref{ass1:smalljumps} and \eqref{ass1:bigjumps} of Assumption \ref{ass:1}. Therefore, the dominated convergence theorem can be used to produce the representation \eqref{rep:varphi}.
The second claim follows similarly. The strict monotonicity property in the last claim follows directly from \eqref{rep:varphi}. Finally, \eqref{ass1:pm} ensures that the map $\varphi_i(\cdot,u^{(i)})$ is onto $\R$. 

$\endproof$

\begin{lemma}\label{lem:OU} Suppose Assumption \ref{ass:1} holds. Let $r\in\R$ be a constant and define $A$ by \eqref{def:A}. For any two continuous functions $(m,c)$ on $[0,T]$ there exists a unique solution of the linear SDE
\begin{align}\label{OUbridge}
dX_t = \Big( \big(r-A(t)^{-1}\big)X_t +m(t) \Big)dt+A(t)c(t) \int_{\R^{I+1}} z^{(0)} \tilde{N}(dt,dz), \quad X_0 \in\R,
\end{align} 
on $[0,T)$ with $X_t \to 0$, $\P$-a.s., as $t\uparrow T$. \end{lemma}
\proof The proof only requires the following modification to the proof of Lemma 1 in \cite{CLM12}:  
For $b(t) = r - A(t)^{-1}$ we claim that the martingale
$$
I_t = \int_0^t \int_{\R^{I+1}}  e^{-\int_0^sb(u)du}A(s)c(s)z^{(0)} \tilde{N}(ds,dz),\quad t\in[0,T),
$$
is uniformly bounded in $\L^2(\P)$. In that case, the martingale convergence theorem ensures that $I_t$ converges $\P$-a.s. to an $\R$-valued random variable as $t\uparrow T$. We can express \eqref{def:A} as
$$
A(t) = \int_0^t e^{-r(s-t)}ds=  \frac1{r}(1-e^{-r(T-t)}),\quad t\in[0,T].
$$ 
Consequently, $\int_0^t A(u)^{-1}du\uparrow +\infty$ as $t\uparrow T$ which implies that $e^{\int_0^tb(u)du}I_t$ converges to zero as $t\uparrow T$.

To see the claimed $\L^2(\P)$-boundedness, we compute for $t\in[0,T)$ the expected quadratic variation
\begin{align*}
\E\Big[ [I]_t\Big] &=  \int_0^t \int_{\R^{I+1}}  e^{-2\int_0^sb(u)du}A(s)^2c(s)^2(z^{(0)})^2\nu(dz)ds\\
&\le T \sup_{s\in[0,T]} e^{-2\int_0^sb(u)du}A(s)^2c(s)^2\int_{\R^{I+1}}(z^{(0)})^2\nu(dz).
\end{align*}
Because all involved functions are continuous, it suffices to observe
$$
\lim_{s\uparrow T} e^{-2\int_0^sb(u)du}A(s)^2c(s)^2=\lim_{s\uparrow T} A(0)^2c(s)^2= A(0)^2 c(T)^2<\infty.
$$
The first equality follows because $\frac{\partial}{\partial t}A(t)^{-2} = -2A(t)^{-2}b(t)$ whereas the continuity of $c$ produces the last equality.

$\endproof$

\section{Proofs}

\proof[Proof of Theorem \ref{thm:investor}] The proof is split into a number of steps:

\noindent Step 1: We start by verifying that \eqref{psii} produces an equivalent martingle measure $\Q^{(i)}$. We can use \eqref{thetaistar} to re-write  \eqref{psii} as
\begin{align}\label{inv1}
\begin{split}
\psi_{i}(z) &= e^{f^i_{-\tfrac{1}{\tau_i}\sigma_i}\left(-\frac{\mu(t)}{\sigma_DA(t)}+\int_{\R^{I+1}} z^{(0)}\nu(dz)\right)z^{(0)} -\tfrac1{\tau_i} \sigma_i z^{(i)}}-1,\quad z\in \R^{I+1}.
\end{split}
\end{align}
To see the integrability property
$$
\int_0^T\int_{\R^{I+1}} \big(\psi_{i}(z)\big)^2\nu(dz)dt<\infty,
$$
it suffices to note that $f^i_{-\tfrac{1}{\tau_i}\sigma_i}(\cdot)$ is continuous and then use the bound $|e^z-1|\le |z|e^{|z|}$, $z\in\R$, together with the integrability conditions in Assumption \ref{ass:1}. Novikov's condition for L\'evy processes (see Theorem 9 in \cite{PS2008}) ensures that the solution $Z^{\psi_i}$ of \eqref{Zpsi} is a strictly positive martingale. The sigma-martingale property \eqref{psi:c3} follows from the representation \eqref{inv1} and the definition of the inverse function $f^i$ (see \ref{fi}).

\noindent Step 2: We next verify the admissibility of $\theta_i^*$ defined by \eqref{thetaistar}  and $c_i^*$ defined by \eqref{cistar}. For $t\in[0,T)$, we insert $(\theta^*_i,c_i^*)$ into the wealth dynamics \eqref{dX} to produce the dynamics
\begin{align*}
dX^*_{it} = \Big(&\big(r-A(t)^{-1}\big) X^*_{it} -\tau_i A(t)^{-1} \int_t^T e^{-r(s-t)} g_i(s-t)ds +\theta_i^*\mu(t)\Big) dt \\& + \theta^*_i\sigma_DA(t) \int_{\R^{I+1}} z^{(0)} \tilde{N}(dt,dz).
\end{align*}
By continuity of all involved functions, these dynamics are well-defined on $[0,T)$. To apply Lemma \ref{lem:OU} we need to show that
\begin{align}\label{def:m}
m(t)= A(t)^{-1} \int_t^T e^{-r(s-t)}g_i(s-t)ds+\theta_i^*\mu(t)
\end{align}
is a continuous function on $[0,T]$. The only potential problem is for the first term when $t=T$. However, L'Hopital's and Leibnitz's rules produce
$$
\lim_{t\uparrow T} \frac{\int_t^T e^{-r(s-t)} g_i(s-t)ds}{A(t)}=\lim_{t\uparrow T} \frac{\int_t^T e^{-r(s-t)} \left(rg_i(s-t)-g_i\right)ds}{rA(t)-1 }=0.
$$
To finish this step we will next prove that  $X^*_{it}/S^{(0)}_t + \int_0^t c^*_{iu}/S^{(0)}_udu$ is a $\Q^{(i)}$-martingale for $t\in[0,T]$. Girsanov's theorem produces the $\Q^{(i)}$-dynamics
$$
d\frac{X^*_{it}}{S^{(0)}_t}+\frac{c^*_{it}}{S^{(0)}_t}dt= \frac{\theta^*_i \sigma_DA(t)}{S_t^{(0)}} \int_{\R^{I+1}} z^{(0)} \Big(N(dt,dz)-\big(1+\psi_{i}(z)\big)\nu(dz)dt\Big).
$$
The needed martingale property follows from 
$$
\int_{\R^{I+1}} (z^{(0)})^2\big(1+\psi_{i}(z)\big)\nu(dz)<\infty.
$$
This integrability property follows from the definition of $\psi_i$ (see \ref{psii}) and the integrability requirements in Assumption \ref{ass:1}.

\noindent Step 3: We need to verify that the pair $(\theta^*_i,c^*_i)$ attains the supremum in  \eqref{primal}. We start by showing that there exists a constant $\alpha>0$ such that
\begin{align}\label{FOCi}
U'_i\big(c^*_{it}+Y_{it}\big) = \alpha \frac{Z^{\psi_i}_t}{S^{(0)}_t},\quad t\in[0,T].
\end{align}
It\^o's lemma produces the dynamics of the left-hand-side to be:
\begin{align*}
\frac{dU'_i\big(c^*_{it}+Y_{it}\big)}{U'_i\big(c^*_{it-}+Y_{it-}\big)}&= \int_{\R^{I+1}} \big(e^{-\tfrac1{\tau_i}\big(\theta^*_i\sigma_Dz^{(0)} +\sigma_iz^{(i)}\big)}-1\big) \tilde{N}(dt,dz)\\&+ \int_{\R^{I+1}} \Big(e^{-\tfrac1{\tau_i}\big(\theta^*_i\sigma_Dz^{(0)} +\sigma_iz^{(i)}\big)}-1+\frac1{\tau_i}\big(\theta^*_i\sigma_Dz^{(0)} +\sigma_iz^{(i)}\big)\Big) \nu(dz)dt\\
& -\frac1{\tau_i}\Big(\theta^*_i \mu(t)A(t)^{-1}-g_i\tau_i +\mu_i \Big)dt\\
&=\int_{\R^{I+1}} \psi_{i}(z) \tilde{N}(dt,dz) -rdt,
\end{align*}
where the last equality uses $g_i$'s definition (see \ref{go}) as well as $\psi_i$'s definition (see \ref{psii}).

We now have all the ingredients needed to perform verification: For any $(\theta,c_i)\in\sA_i$ we have
\begin{align*}
\E\left[\int_0^T U_i (c_{it} +Y_{it}) dt\right] &\le \E\left[\int_0^T V_i(\alpha \frac{Z^{\psi_i}_t}{S^{(0)}_t})dt +\alpha \int_0^T \frac{Z^{\psi_i}_t}{S^{(0)}_t} (c_{it} +Y_{it}) dt+\alpha Z^{\psi_i}_T\frac{X_{iT}}{S^{(0)}_T} \right]\\
&\le \E\left[\int_0^T V_i(\alpha \frac{Z^{\psi_i}_t}{S^{(0)}_t})dt +\alpha \int_0^T \frac{Z^{\psi_i}_t}{S^{(0)}_t} Y_{it} dt\right]+\alpha X_{i0} \\
&= \E\left[\int_0^T V_i(\alpha \frac{Z^{\psi_i}_t}{S^{(0)}_t})dt +\alpha \int_0^T \frac{Z^{\psi_i}_t}{S^{(0)}_t} (c^*_{it} +Y_{it}) dt \right]\\
&=\E\left[\int_0^T U_i (c^*_{it} +Y_{it}) dt\right].
\end{align*}
The function $V_i$ is the Fenchel conjugate $V_i(y) = \sup_{x>0} \big( U_i(x) -xy\big), \;y>0$. The first inequality uses Fenchel's inequality and the non-negativity of $\alpha$ and $X_{iT}$. The second inequality uses the $\Q^{(i)}$-supermartingale property of admissible controls. The first equality uses the martingale property proven in Step 2 and $X^*_{iT} =0$.  Lemma 3.4.3(iv) in \cite{KS98} proves the relation $U(x) = V\big(U'(x)\big)+xU'(x)$ which provides the last equality.

$\endproof$

\proof[Proof of Theorem \ref{thm:eq}] We start by noticing that the requirement \eqref{eq:mu} ensures that $\mu A^{-1}$ is constant. Because $A$ defined by \eqref{def:A} is continuous, we see that $\mu$ is also continuous on $[0,T]$.  Consequently, the pair $(r,\mu)$ satisfies Assumption \ref{ass2}.

For $\psi^*$ defined by \eqref{psi*} the corresponding exponential \eqref{Zpsi} is a strictly positive martingale. This follows from 
\begin{align}\label{integrability1}
\int_0^T\int_{\R^{I+1}} \big(\psi_t^*(z)\big)^2\nu(dz)dt<\infty,
\end{align}
and Novikov's condition for L\'evy processes (see Theorem 9 in \cite{PS2008}). The integrability property \eqref{integrability1} follows from the definition of $\psi^*$ (see \ref{psi*}) and the integrability requirements in Assumption \ref{ass:1}.
Consequently, $\Q^*$ is a well-defined equivalent martingale measure. Girsvanov's theorem and \eqref{psi*} change the $\P$-dynamics \eqref{dD} to the following $\Q^*$-dynamics
$$
dD_t = \Big(\mu_D - \mu(t)A(t)^{-1}\Big)dt + \sigma_D\int_{\R^{I+1}} z^{(0)} \Big(N(dt,dz)-\big(1+\psi_t^*(z)\big)\nu(dz)dt\Big).
$$
To ensure that the stochastic integral is a $\Q^*$-martingale it suffices to see
$$
\int_{\R^{I+1}} (z^{(0)})^2\big(1+\psi_t^*(z)\big)\nu(dz)<\infty.
$$
This follows from $\psi^*$'s definition \eqref{psi*} and the integrability requirements in Assumption \ref{ass:1}. Fubini's Theorem for conditional expectations produces
\begin{align}
\begin{split}
S_t &=\int_t^T e^{-r(s-t)du}  \E^{\Q^*}[D_s|\sF_t]ds \\
&=A(t)D_t+\int_t^T e^{-r(s-t)du}\int_t^s \big(\mu_D - \mu(u)A(u)^{-1}du\big)ds.\label{eq1}
\end{split}
\end{align}
This representation and the property $A'(t) = rA(t) -1$ produce the dynamics \eqref{dS}. To see that $S_{T-}=0$ we can use \eqref{eq1} to re-write the dynamics \eqref{dS} as
\begin{align}
dS_t&= \Big( \big(r - A(t)^{-1}\big)S_t +\mu(t) +A(t)^{-1}S_t - D_t\Big)dt +\sigma_DA(t)\int_{\R^{I+1}}  z^{(0)} \tilde{N}(dt,dz)\nonumber\\
&= \Big(A(t)^{-1}\int_t^T e^{-r(s-t)du}\int_t^s \big(\mu_D - \mu(u)A(u)^{-1}du\big)ds\label{eq2}\\&+ \big(r - A(t)^{-1}\big)S_t +\mu(t) \Big)dt +\sigma_DA(t)\int_{\R^{I+1}} z^{(0)} \tilde{N}(dt,dz).\nonumber
\end{align}
 The deterministic part of the drift in \eqref{eq2}, i.e., 
$$
m(t)= A(t)^{-1}\int_t^T e^{-r(s-t)du}\int_t^s \big(\mu_D - \mu(u)A(u)^{-1}du\big)ds +\mu(t),
$$
is a continuous function on $[0,T]$. This can be seen similarly as the argument following \eqref{def:m} above. Lemma \ref{lem:OU} then produces $S_{T-}=0$.

To see that the clearing conditions \eqref{clearingconds} hold, we first note that \eqref{eq:mu} ensures that the stock market clears, i.e., $\sum_{i=1}^I \theta^*_i =1$. Clearing for the money market account market is equivalent to $S_t = \sum_{i=1}^I X^*_{it}$. For $t=0$ this holds. 
We can use \eqref{cistar} and $\sum_{i=1}^I \theta_i^*(t)=1$ to find the following dynamics 
\begin{align*}
\sum_{i=1}^I dX^*_{it} &=\Big( \big(r-A(t)^{-1}\big)\sum_{i=1}^I X^*_{it}-\sum_{i=1}^I\tau_i A(t)^{-1} \int_t^T e^{-r(s-t)}g_i(s-t)ds+\mu(t) \Big)dt \\&+ \sigma_DA(t) \int_{\R^{I+1}} z^{(0)} \tilde{N}(dt,dz).
\end{align*}
Therefore, the representation \eqref{eq2} produces the equivalent requirement
$$
-\sum_{i=1}^I \tau_i g_i = \mu_D - \mu(t)A^{-1}(t),\quad t\in[0,T].
$$
To see that this relationship holds we can insert the definition of $g_i$ (see \ref{go}) and use the definition of $r$ (see \ref{eq:r}). We note that this argument also produces clearing in the good's market, i.e., $\sum_{i=1}^I c_{it}^* = D_t$, which finishes the proof.

$\endproof$

\end{document}